\documentclass[modern]{aastex61}

\usepackage{graphicx}
\usepackage{amsmath}
\usepackage{amssymb}
\usepackage{enumitem}

\newcommand\mybar{\kern1pt\rule[-\dp\strutbox]{.8pt}{\baselineskip}\kern1pt}

\setlist[itemize]{noitemsep, topsep=0pt, leftmargin=*}

\shorttitle{Gravitational Redshift in Wide Binaries}
\shortauthors{Loeb}

\begin{document}

\title{Gravitational Redshift for Wide Binaries in {\it Gaia} eDR3}

\author{Abraham Loeb}
\affiliation{Astronomy Department, Harvard University, 60 Garden
  St., Cambridge, MA 02138, USA}

\begin{abstract}
The Doppler effect is commonly used to infer the velocity difference
between stars based on the relative shifts in the rest-frame
wavelengths of their spectral features. In wide binaries, the
difference in gravitational redshift from the surfaces of the
constituent stars with distinct compactness dominates at separations
$\gtrsim 10^{-2}$ pc.  I suggest that this effect became apparent for
wide pairs in the {\it Gaia} eDR3 catalogue but was incorrectly
interpreted as a possible modification of Newtonian gravity in the
internal kinematics of very wide binaries.
\end{abstract}

\section{Introduction}

According to General Relativity, the spectrum of radiation emitted
from the surface of a star is gravitationally redshifted relative to a
distant observer.  In the weak field regime, the spectroscopic
velocity shift $v_{\rm gr}$ is given by \citep{1972gcpa.book.....W},
\begin{equation}
v_{\rm gr}=-{GM\over c R} = - 0.636 \left({M\over M_\odot}\right)
\left({R\over R_\odot}\right)^{-1}~{\rm km~s^{-1}} ,
\label{one}
\end{equation}
for a star of mass $M$ and photospheric radius $R$.  The observed
radii of stars show significant scatter for masses $\gtrsim 1 M_\odot$
due to stellar evolution and rotation \citep{2010A&ARv..18...67T},
implying that redshift variations of $\sim 0.5~{\rm km~s^{-1}}$ should
be common in pairs of stars with different compactnesses. Attempts to
measure the gravitational redshift in main-sequence stars are
challenging \citep{2011A&A...526A.127P}, but recent data indicated
otherwise \citep{2021arXiv210201079M}. Fluctuations as a result of surface
turbulence would average out for a large enough statistical sample of
stars. A larger signal of gravitational redshift had already been
detected for compact stars, such as white dwarfs
\citep{1967AJ.....72Q.301G,2005MNRAS.362.1134B,2012ApJ...757..116F}
and neutron stars \citep{2002Natur.420...51C}.

The mass-radius relation for main-sequence stars of up to a few solar
masses follows an approaximate power-law relation
\citep{2011AdAst2011E..13R}, $(R/R_\odot)\approx
(M/M_\odot)^{0.8}$, implying,
\begin{equation}
v_{\rm gr} \approx - 0.6 \left({M\over M_\odot}\right)^{0.2}~{\rm
  km~s^{-1}} .
\label{three}
\end{equation}

\section{Implications} 

The circular velocity $(GM/r)^{1/2}$ of a test particle around a star
of mass $M$ falls below the velocity shift $v_{\rm gr}$ at orbital
radii $r$ that exceed a critical value \citep{2014PhRvD..89b7301L},
\begin{equation}
r_{\rm crit} \equiv \left({c^2 R^2\over GM}\right)= 10^{-2}
\left({M\over M_\odot}\right)^{0.6}~{\rm
pc} ,
\label{two}
\end{equation}
Hence, the spectroscopic velocity difference between the stellar
members of a wide binary separated by more than $\sim 10^{-2}~{\rm
  pc}=2\times 10^3~{\rm au}$ could be significantly affected by the
difference in their intrinsic gravitational redshifts owing to their
distinct compactnesses. At separations $r\gtrsim r_{\rm crit}$, the
actual radial velocity difference could be smaller than the
gravitational redshift difference, implying a constant Doppler shift
{\it independent of separation}. Wide binaries at separations $\gtrsim
10^{-2}$~pc are characterized by long orbital times $\gtrsim 10^5$~yr
and low speeds $\lesssim 0.7~{\rm km~s^{-1}}$, making it challenging
to measure any temporal change in the radial velocities of member
stars.

Eccentric binaries spend a substantial fraction of their orbital time
near apocenter where their relative speed is small and the fractional
impact of the gravitational redshift effect is maximized.  At
separations larger than $2r_{\rm crit}$, a binary in a radial velocity
catalogue may be incorrectly declared as unbound if the
gravitational redshift differential is not corrected for.

At the same time, selection for a small differences in radial
velocities between stars at wide separations on the sky inevitably
leads to a selection bias for {\it gravitationally-unbound pairs}
which are observed to have a small spectroscopic velocity difference
between their members just because their actual radial velocity
difference happens to be compensated by the difference in
gravitational redshift due to their distinct compactnesses. Being
gravitationally unbound, these wide pairs with small spectroscopic
shifts, would have proper motions on the sky that are
larger than expected for gravitationally bound systems. On average in
a large statistical sample, the excess proper motion in the sky would
be of the same magnitude as the gravitational redshift - given a
random geometric orientation of the velocity vector for each star.

\section{{\it Gaia} eDR3 Catalogue}

Recently, the kinematics of wide pairs of main-sequence stars with a
total mass of $(1$--$2.2)M_\odot$ in the {\it Gaia} eDR3 catalogue was
selected with strict data cuts on signal-to-noise ratio and radial
velocities \citep{2022MNRAS.509.2304H}. For separations below
$0.009$~pc, the results showed the expected Newtonian behavior of the
velocity difference between the binary members scaling in proportion
to $(M_b/r)^{1/2}$, where $M_b$ is the total binary mass. But at
separations $\gtrsim 0.009$~pc, the data showed a
separation-independent velocity difference of $\sim 0.5~{\rm
  km~s^{-1}}$ that scales weakly with binary mass, $\propto
M_b^{0.24\pm0.21}$. The authors suggested that this situation is
reminiscent of the deviation from Newtonian gravity in the baryonic
Tully-Fisher relation \citep{2020Galax...8...35M}, where at low
accelerations there is a need for either dark matter or modified
gravity. But they also noted that the results are at odds with the
predictions of Modified Newtonian Dynamics, abbreviated as MOND
\citep{2020SHPMP..71..170M}, where the external field effect implies
only small deviations from Newtonian dynamics.

\section{Conclusions} 

Here I suggest that the above results reflect the selection bias
induced by gravitational redshift to radial velocity cuts of
spurious unbound pairs that were incorrectly identified as being
gravitationally bound. This interpretation is consistent with three
characteristics of the reported results: {\it (i)} the agreement
between the measured asymptotic velocity of $\sim 0.5~{\rm km~s^{-1}}$
and equation (\ref{one}); {\it (ii)} the weak dependence of the
asymptotic velocity at large separations on the total pair mass,
$\propto M_b^{0.2}$, as expected from equation (\ref{three}); and {\it
  (iii)} the agreement between the transition separation of $0.009$~pc
and the value of $r_{\rm crit}$ in equation (\ref{two}).

High resolution spectra of stars can be used to infer their surface
gravity, $GM/R^2$, and correct for the above-mentioned selection bias.

\bigskip
\bigskip
\section*{Acknowledgements}

This work was supported in part by Harvard's {\it Black Hole
  Initiative}, which is funded by grants from JFT and GBMF. I thank
Morgan MacLeod for comments on the manuscript.

\bigskip
\bigskip
\bigskip

\bibliographystyle{aasjournal}
\bibliography{g}
\label{lastpage}
\end{document}